\documentclass[aps,twocolumn,pra,superscriptaddress,showpacs,tightenlines]{revtex4}
\usepackage{graphicx,amsmath,amsfonts,amssymb}
\usepackage{times}
\begin{document}

\title{Entanglement-enhanced two-photon delocalization in a coupled-cavity array}
\author{Shi-Qing Tang, Ji-Bing Yuan, Xin-Wen Wang, and Le-Man Kuang\footnote{Author to whom any correspondence should be
addressed. }\footnote{ Email: lmkuang@hunnu.edu.cn}}

\affiliation{Key Laboratory of Low-Dimensional Quantum Structures
and Quantum Control of Ministry of Education,  and Department of
Physics, Hunan Normal University, Changsha 410081, China}

\begin{abstract}
We study transport properties of two entangled photons which are
initially injected into two  nearest-neighbor coupling cavities in a
one-dimensional coupled-cavity array (CCA). It is found that
photonic transport dynamics in the two-photon CCA exhibits
entanglement-enhanced two-photon delocalization (TPD) phenomenon. It
is shown that the CCA can realize the localization-to-delocalization
transition for two entangled photons.
\end{abstract}

\pacs{03.67.-a, 42.50.Pq, 37.30.+i}
\keywords{two-photon
delocalization, a coupled-cavity array, quantum entanglement}

\maketitle

\section{Introduction}
In recent years, much attention has been paid to coupled-cavity
arrays (CCAs) \cite{ill,gree,hart1,hart2,hart3,tom,hou,sch,mlm,lom}
due to their extremely rich physics as well as their wide potential
applications to work as an effective platform to realize quantum
information processing and to carry out photonics tasks. For
instance, CCAs stand out as an attractive controllable test bed for
strongly correlated many-body models and quantum simulators of
many-body physics \cite{ill,gree,hart1,hart2,hart3,gre,ang}.
Transport properties of photons and various excitations  in CCAs
embedded with natural or artificial atoms are also receiving
considerable interest
\cite{zhou1,zhou2,zhou3,ogd,mak,bose,qua,pat,lon,gon,shi,liao}. In
particular, it has been demonstrated that  CCAs with a cyclic
three-level system or a single atom open the possibility to quantum
routing of single photons \cite{zhou4,lu}.

CCAs typically consist of an arrangement of low-loss cavities with
nearest-neighbor coupling allowing photon hopping between
neighboring cavities.  An appealing feature of CCAs is the available
high control of each single cavity in terms of performable
measurements, quantum-state engineering, and dynamical parameter
tuning. This feature can be used to realize local addressing in
quantum information processing. Progress in the fabrication
techniques make CCAs experimentally accessible \cite{wal,hen,far}.
Although quantum entanglement is an essential resource in quantum
information processing, propagation of spatially entangled photons
in CCAs without natural or artificial atoms has not been studied
extensively heretofore. A natural question is how quantum
entanglement of initial input photons affects photonic transport
properties in CCAs.

In this paper, we  study transport properties of two initially
entangled photons which are injected into two nearest-neighbor
coupling cavities in a one-dimensional CCA. We show that photonic
transport dynamics in the two-photon CCA exhibits the
entanglement-enhanced two-photon delocalization (TPD) phenomenon. We
investigate TPD dynamics by introducing the concept of the TPD
degree when two nearest-neighbor coupling cavities are initially in
two-photon entangled state $|\Psi\rangle=\sin\theta |2,0\rangle+
\cos\theta |0, 2\rangle$. We demonstrate that the CCA can realize
the localization-to-delocalization transition for two entangled
photons.

The rest of this paper is organized as follows. In Sec II, we
present our physical model and its  solution. In Sec III, we analyze
photonic transport properties and TPD in the two-photon CCA, and
show the entanglement-enhanced TPD and the
localization-to-delocalization transition for two entangled photons.
Finally, we shall conclude our paper with discussions and remarks in
the last section.

\section{Physical model and solution}

Let us consider a finite-length CCA in which nearest-neighbor
cavities are coupled together so as to allow for photon hopping. We
assume that  all the cavities are identical. Each cavity sustains a
single field mode of frequency $\omega$.  The Hamiltonian of the
full system (here and throughout we take $\hbar = 1$) is written as
\begin{equation}
\hat{H}=\omega \sum\limits_{j=1}^{N}a_{j}^{\dag
}a_{j}+J\sum\limits_{j=1}^{N-1}\left( a_{j}^{\dag
}a_{j+1}+H.c.\right), \label{1}
\end{equation}
where $\hat{a}^{\dag} (\hat{a})$ is a photonic creation
(annihilation) operator at $i$-th cavity. The second term in above
equation represents the hopping interaction with the intercavity
coupling strength $J$.

The Hamiltonian of the CCA given by Eq. (1) can be diagonalized in
terms of the normal modes defined by
\begin{equation}
\hat{c}_{k}(t)=\sum\limits_{j=1}^{N}\hat{a}_{j}(t)S\left(j,k\right),
\label{2}
\end{equation}
with the following inverse transformation
\begin{equation}
\hat{a}_{j}(t)=\sum\limits_{k=1}^{N}\hat{c}_{k}(t)S\left(j,k\right),
\label{3}
\end{equation}
where the transformation matrix is given by
\begin{equation}
S\left(j,k\right)=\sqrt{\frac{2}{N+1}}\sin\left(\frac{j\pi
k}{N+1}\right).
 \label{4}
\end{equation}

Making use of Eqs. (2)-(4), one can obtain the diagonalized form of
Hamiltonian (1) where the transformation matrix is given by
\begin{equation}
\hat{H} =\sum\limits_{k=1}^{N}\Omega_{k}c_{k}^{\dag }c_{k},
\hspace{0.3cm}  \Omega_{k}=\omega +2J\cos \left(\frac{\pi
k}{N+1}\right). \label{5}
\end{equation}

From Heisenberg equations on Eq. (5) and making use of Eq. (3), one
can obtain exact solution of the CCA in the heisenberg
representation with the following form
\begin{equation}
a_{j}(t)=\sum_{l}G_{jl}(t)a_{l}(0). \label{6}
\end{equation}
where the explicit form of the Green's function is given by
\begin{equation}
G_{j,l}(t)=\sum\limits_{k=1}^{N}\exp\left\{-i \Omega _{k}
t\right\}S\left(j,k\right)S\left(l,k\right). \label{7}
\end{equation}

In the following section, we shall use above exact solution to
investigate quantum dynamics of photonic transport and TPD in the
two-photon CCA.

\section{Entanglement-enhanced two-photon
delocalization}

The quantum mechanical properties of light are observed when
correlations between the propagating photons are considered. In this
section, we study photonic transport properties and quantum dynamics
of  two-photon delocalization in the two-photon CCA when two
entangled photons are injected into the CCA by investigating the
evolution of the photon-number correlation function
\begin{equation}
P_{m,n}(t)=\left\langle a_{n}^{\dag }(t)a_{m}^{\dag
}(t)a_{m}(t)a_{n}(t)\right\rangle, \label{8}
\end{equation}
which represents the probability of detecting a two-photon
coincidence across cavity $m$ and cavity $n$. The probability to
detect both photons at the same cavity $n$ is given by
$P_{n,n}(t)/2$.

We consider such a situation in which the two photons are initially
in $r$-th cavity and  $s$-th cavity with the following two-photon
NOON-type states
\begin{equation}
\left\vert \protect\psi \right\rangle  = \sin\theta\left\vert
2\right\rangle _{r}\left\vert 0\right\rangle _{s} +
\cos\theta\left\vert 0\right\rangle _{r}\left\vert 2\right\rangle
_{s}, \label{9}
\end{equation}
where for the simplicity, we take $0\leqslant\theta \leqslant\pi/2$
throughout the paper. It is interesting to note that the two photons
in the two-photon entangled states (9) are in the same cavity, i.e.,
cavity $r$ or cavity $s$. The other cavities in the CCA are in
vacuum state. Hence, the two-photon entangled states  given by Eq.
(9) are two-photon localization states. The amount of entanglement
of the two-photon states in Eq.(9) can be measured by concurrence
\cite{woo} with the following expression
\begin{equation}
\mathcal{C}(\theta)=|\sin 2\theta|. \label{10}
\end{equation}
Obviously, the concurrence vanishes when $\theta=0$, and $\pi/2 $
while the concurrence reaches the maximum value $1$ when
$\theta=\pi/4 $.

Substituting Eqs. (6) and (9) into Eq. (8), we can obtain the
probability of detecting a two-photon coincidence across cavity $m$
and cavity $n$ with the following expression
\begin{equation}
P_{mn} =2\left\vert \cos \theta G_{mr}(t)G_{nr}(t)+\sin \theta
G_{ms}(t)G_{ns}(t)\right\vert ^{2}.
\end{equation}

As an example, we consider the photonic propagation in the CCA for
the two-photon entangled state given by Eq. (9) which is injected
initially into two neighboring cavities, say $r=15$ and $s=16$.
Assume that the CCA under our consideration consists of 29 cavities
($N=29$). Quantum evolution of the calculated correlation function
for this case is presented in Fig. 1 for different amount of initial
entanglement. Fig.1(a), Fig.1(c), and Fig.1(e) are the initial-state
correlation function when the amount of initial entanglement
$\mathcal{C}=0, 0.5,$ and $1$, respectively. Fig.1(b), Fig.1(d), and
Fig.1(f) are the correlation function at time $\omega t=83.57$ when
the amount of initial entanglement $\mathcal{C}=0, 0.5,$ and $1$,
respectively. From Fig. 1 we can see that the correlation
represented by  the diagonal elements of the two-photon correlation
function $P_{mn}$ completely vanishes while the correlation
represented by the off-diagonal elements of the two-photon
correlation function $P_{mn}$ is enhanced  with the increase of the
initial amount of entanglement from vanishing entanglement
($\mathcal{C}=0$) to the maximal entanglement ( $\mathcal{C}=1$).
Accordingly, the two photons will always tend to separate and to
emerge in different cavities.

\begin{figure}
  \centering
  \includegraphics[height=10cm, width=3.3in]{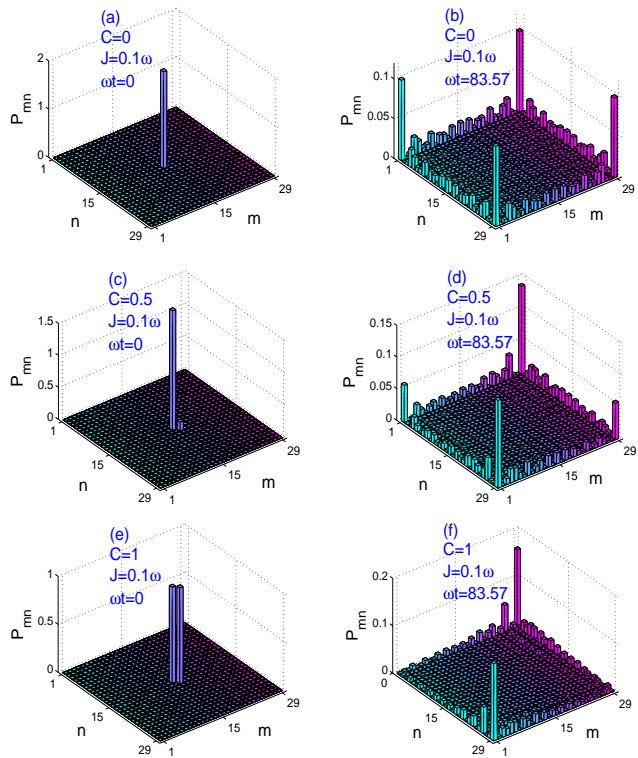}
  \caption{(Color online) The joint probability $P_{mn}$ of simultaneously
detecting photons in cavity $m$ and $n$ after a temporal evolution
$\omega t=83.57$. The evolution time is  scaled by  $1/J$, and other
parameters are $\omega=J$, $N=29$.}
\end{figure}

In order to well understand transport dynamics of the two photons in
the CCA, we introduce the concept of the two-photon delocalization
(TPD). When the two photons are in two different cavities in the
two-photon CCA we call it as the TPD while when the two photons are
in the same cavity we call it as the two-photon localization (TPL).
The degree of the TPD can be defined in terms of the sum of
probability $P_{nn}/2$ of detecting photons in every cavity
\begin{equation}
\eta=1-\frac{1}{2}\sum\limits_{n=1}^{N} P_{n,n}(t), \label{12}
\end{equation}
which represents the probability of the two photons being not at the
same cavity in the CCA simultaneously. Especially, when $\eta=1$,
the two photons are in the complete TPD state in which the two
photons are completely in different cavities. In this case, it is
impossible that the two photons are in the same cavity. Hence, the
two photons in the TPD state tend to stay at different cavities. In
this sense, the TPD degree can reflect repulsive-statistics or
anti-bunching property of two photons in the CCA. Obviously, the TPD
degree of the two-photon NOON-type state defined by Eq. (9) is zero.
Hence, the state given by Eq. (9) is not a TPD state, it is a TPL
state. The larger the value of the parameter $\eta$ is, the stronger
the TPD phenomenon becomes. We consider the two-photon CCA formed by
29 cavities. When the two photons are injected into $15$-th and
$16$-th cavities with the input state (9), the photon-number
second-order correlation function changes considerably in the time
evolution as shown in Fig. 1. The most obvious feature is the
appearance of the TPD in which the diagonal elements of the
two-photon probability, i.e., $P_{nn}/2$, are suppressed
considerably  in the time evolution. In particular, for the case of
the maximally entangled state with the concurrence $\mathcal{C}=1$
we can observe almost completeTPD with $P_{nn}\approx 0$ in the
process of the time evolution as shown in Fig. 1(e) and Fig. 1(f).

\begin{figure}[tbp]
\centering
  \includegraphics[height=6cm, width=8.5cm]{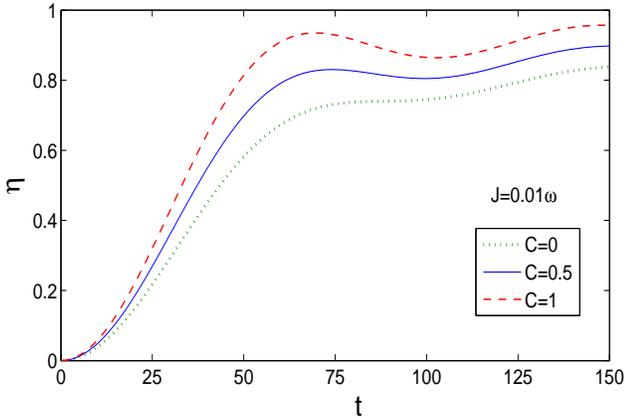}
\caption{(Color online)  Time evolution of the two-photon
delocalization degree $\eta$  in the weak hopping regime  for three
different values of quantum entanglement of the two-photon state
given by Eq. (9). The evolution time is  scaled by  $1/J$. The
photon hopping parameter takes $J=0.01\omega$. The number of the
cavities in the CCA is $N=29$. The  dashed curve represents the
result of the maximally entangled state with $\mathcal{C}=1$, the
solid line corresponds to the initial input state with the
concurrence $\mathcal{C}=1$, and the dot line shows the result of
the initial unentangled state with $\mathcal{C}=0$. }
\label{fig2.eps}
\end{figure}

We now investigate quantum dynamics of the TPD and the influence of
quantum entanglement of the initial state in the weak hopping regime
of $J\ll \omega$. In  Fig. 2, we have plotted the time evolution of
the TPD degree for three different values of quantum entanglement of
the initial state (9), $\mathcal{C}=, 0, 0.5$, and $1$,
respectively. We here take the photon hopping strength
$J=0.01\omega$ and the number of the cavities in the CCA  $N=29$.
The dashed curve represents the case of the maximally entangled
state with $\mathcal{C}=1$, the solid line corresponds to the
initial input state with the concurrence $\mathcal{C}=0.5$, and the
dot line represents the case of the initial unentangled state with
$\mathcal{C}=0$. From Fig. 2 we can see that the two photons in the
CCA rapidly tend to the TPD with the time evolution in a short-time
regime. This clearly indicates the transition from the TPL state to
the TPD state in the two-photon CCA. The TPD degree can be preserved
in the process of the long time evolution afterwards except some
small oscillations. On the other hand, Fig.2 indicates that the TPD
is enhanced with the increase of quantum entanglement of the initial
state (9). Hence, we can conclude that quantum entanglement can
essentially affect the TPD.

\begin{figure}[tbp]
\centering
  \includegraphics[height=6cm, width=8.5cm]{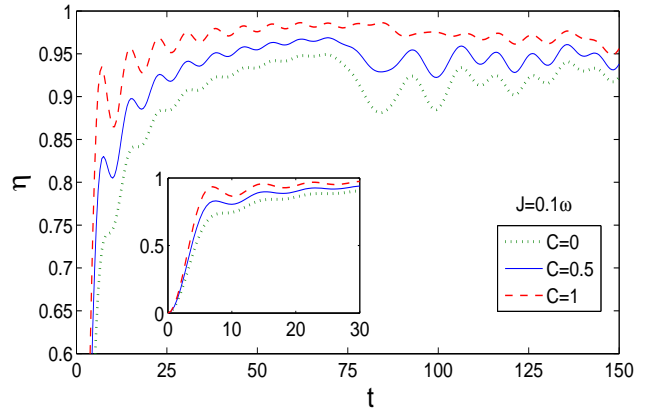}
\caption{(Color online)  Time evolution of the two-photon
delocalization degree $\eta$  in the strong hopping regime for three
different values of quantum entanglement of the two-photon state
given by Eq. (9). The evolution time is  scaled by  $1/J$. The
photon hopping parameter takes $J=0.1\omega$. The number of the
cavities in the CCA is $N=29$. The  dashed curve represents the
result of the maximally entangled state with $\mathcal{C}=1$, the
solid line corresponds to the initial input state with the
concurrence $\mathcal{C}=1$, and the dot line shows the result of
the initial unentangled state with $\mathcal{C}=0$. }
\label{fig2.eps}
\end{figure}

We then turn to  quantum dynamics of the TPD in the strong hopping
regime of $J\approx \omega$. In. Fig. 2, we have plotted the time
evolution of the TPD degree $\eta$ for three different values of
quantum entanglement of the initial state $|\psi\rangle$. The photon
hopping parameter takes $J=0.1\omega$. The number of the cavities in
the CCA is $N=29$. The dashed curve represents the case of the
maximally entangled state with $\mathcal{C}=1$, the solid line
corresponds to the initial input state with the concurrence
$\mathcal{C}=1$, and the dot line represents the case of the initial
unentangled state with $\mathcal{C}=0$. From Fig. 3 we can clearly
observe the entanglement-enhanced TPD phenomenon. In comparison with
the quantum dynamics of the TPD in the weak hopping regime of $J\ll
\omega$,  from Fig.2  and Fig. 3 we can see that the TPD in the
strong hopping regime becomes faster than that in the weak hopping
regime in the short time regime. The oscillations of the TPD degree
in the process of  the time evolution become stronger with the
increasing the photon hopping strength. However, the TPD degree is
over 90 percent in the most time. The short time behavior of the TPD
degree is indicated in the inset in Fig. 3, from which we can see
that the transition from the TPL to the TPD happens in  shorter time
in the strong hopping regime than the weak hopping regime.

\section{Concluding remarks}
In conclusion, we have studied photonic transport properties  and
TPD dynamics in the two-photon CCA when two entangled photons are
initially injected into two nearest-neighbor coupling cavities in
the form of two-photon NOON-type states which are superposition
states of two-photon states $|2,0\rangle$ and $|0,2\rangle$. We have
investigated the TPD dynamics by introducing the concept of the TPD
degree for the initial two-photon entangled states $|\psi\rangle=
\sin\theta |2,0\rangle + \cos\theta |0, 2\rangle $. It has been
found that photonic transport dynamics in the two-photon CCA
exhibits the entanglement-enhanced TPD phenomenon. It has
demonstrated that the CCA can realize the
localization-to-delocalization transition for two entangled photons.
It is worthwhile to point out that the entanglement-enhanced TPD
phenomenon is a two-photon cooperative effect induced by quantum
interference of two entangled photons due to the initial quantum
entanglement. The entanglement-enhanced TPD vanishes in the absence
of entanglement between two photons.

\begin{acknowledgements}
This work was supported by the National 973 Program under Grant No.
2013CB921804 and the NSF under Grant No. 11375060 and Grant No.
11434011.
\end{acknowledgements}

\end{document}